\documentclass[twocolumn,pre,superscriptaddress,aps,footinbib]{revtex4}
\usepackage[usenames,dvipsnames]{color}
\usepackage{url,psfrag,graphicx}
\usepackage{dcolumn}
\usepackage{amsmath,amssymb}
\usepackage{bm}
\usepackage{hyperref}
\usepackage{float,epsfig,color}
\usepackage{times}
\usepackage[utf8]{inputenc}

\usepackage{graphicx}
\usepackage{amsmath}
\usepackage{amssymb}
\usepackage{dcolumn}

\begin{document}

\title{Anomalous dynamic scaling of Ising interfaces}

\author{Enrique Rodr\'{\i}guez-Fern\'andez}
\thanks{Present address: Instituto de F\'{\i}sica de Cantabria (IFCA), Consejo Superior de Investigaciones Cient\'{\i}ficas-Universidad de Cantabria, Avenida de los Castros s/n, 39005 Santander, Spain}
\email{enrodrig@math.uc3m.es}
\affiliation{Departamento de Matem\'aticas and Grupo Interdisciplinar de Sistemas Complejos (GISC)\\ Universidad Carlos III de Madrid, Avenida de la Universidad 30, 28911 Legan\'es, Spain}
\author{Silvia N.\ Santalla}
\email{silvia.santalla@uc3m.es}
\affiliation{Departamento de F\'{\i}sica and GISC, Universidad Carlos III de Madrid, Avenida de la Universidad 30, 28911 Legan\'es, Spain}
\author{Mario Castro}
\email{marioc@iit.comillas.edu}
\affiliation{Instituto de Investigaci\'on Tecnol\'ogica (IIT) and GISC, Universidad Pontificia Comillas, 28015 Madrid, Spain}
\author{Rodolfo Cuerno}
\email{cuerno@math.uc3m.es}
\affiliation{Departamento de Matem\'aticas and Grupo Interdisciplinar de Sistemas Complejos (GISC)\\ Universidad Carlos III de Madrid, Avenida de la Universidad 30, 28911 Legan\'es, Spain}

\begin{abstract}
Until very recently, the asymptotic occurrence of intrinsic anomalous scaling has been expected to require concomitant effects for kinetically rough interfaces, like quenched disorder or morphological instabilities. However, counterexamples have been recently reported for simpler situations dominated by time-dependent noise, as in the discrete growth system associated with an Ising model proposed by H.\ Dashti-Naserabadi {\em et al.}\ [Phys.\ Rev.\ E {\bf 100}, 060101(R) (2019)], who assessed the equilibrium behavior of the model. Here we revisit this system to characterize its time-dependent behavior in two and three dimensions (one-and two-dimensional interfaces, respectively). While the 3D case seems dominated by a fast evolution beyond critical dynamics, in the 2D case numerical simulations of an associated time-dependent Ginzburg-Landau equation retrieve the same static (roughness) exponents and the same intrinsic anomalous scaling Ansatz as in the equilibrium case, throughout the full time evolution. However, the dynamic exponent is seen to cross over between two different values, none of which enables identification with previously known universality classes of kinetic roughening. Simulations for larger system sizes moreover suggest breakdown of scaling behavior at the largest scales, suggesting that the previously reported scaling behavior may be effective and restricted to relatively small systems.
\end{abstract}

\maketitle

\section{Introduction}
\label{sec:intro}

Many spatially-extended systems of a high current interest operate under non-equilibrium conditions, from active \cite{Ramaswamy10,Marchetti13} to quantum matter \cite{Polkovnikov11,Weimer21}. In these and many other contexts, the conditions for and the properties of the emergence of the strong correlations associated with space-time criticality \cite{Tauber14} become relevant. A particularly interesting class of systems from this point of view is that in which critical behavior appears spontaneously without the need for parameter tuning, thus showing so-called generic scale invariance (GSI) \cite{Grinstein95,Belitz05}.

Surface kinetic roughening \cite{Barabasi95,Krug97} is a celebrated physical instance of GSI. Indeed, in many different systems, from thin film production to bacterial colonies, the fluctuations of rough surfaces and interfaces are observed to evolve in the absence of typical scales in time and space. To date, some of the main universality classes of surface kinetic roughening, like that of the Kardar-Parisi-Zhang (KPZ) equation \cite{Kardar86}, are held as paradigms of nonequilibrium critical phenomena at large \cite{Kriechebauer10,Corwin12,Halpin-Healy15,Takeuchi18}, being suprisingly relevant even in contexts far away from those that motivated their original formulation. An example is the recent experimental observation of KPZ scaling for quantum condensates \cite{Fontaine22} and spin chains \cite{Wei22}, the equation being originally formulated for non-quantum systems \cite{Barabasi95,Krug97,Kardar86}.

In general, the study of surface kinetic roughening remains instrumental to generalize the concepts and tools of equilibrium critical dynamics to systems which are far from equilibrium  \cite{Kriechebauer10,Corwin12,Halpin-Healy15,Takeuchi18}. A crucial ingredient at this has been the dynamic scaling ansatz satisfied by the relevant physical quantities, such as the surface roughness and correlations. As seminally proposed by Family and Vicsek (FV) \cite{Family85}, it is a direct generalization of the dynamic scaling ansatz of equilibrium critical dynamics found in e.g.\ the classic models A and B \cite{Hohenberg77,Tauber14}. Indeed, the so-called FV ansatz is satisfied e.g.\ by the KPZ equation and by many other systems displaying kinetic roughening \cite{Barabasi95,Krug97}. However, generalizations of the FV ansatz, collectively termed anomalous scaling \cite{Schroeder93,DasSarma94,Krug97,Lopez97,Ramasco00}, have been later shown to be required to account for more elaborate scaling behavior found in models and in experiments, whereby e.g.\ different critical exponents characterize fluctuations at distances smaller than or comparable to the system size.

Up to quite recently, one of these generalized ansatzs, the so-called intrinsic anomalous scaling, was believed (following a conjecture \cite{Lopez05} based on renormalization group arguments) to occur asymptotically for some rough interfaces only when subject to relatively peculiar conditions, like quenched disorder and/or morphological instabilities. Hence, the observation of intrinsically anomalous scaling in the equilibrium state of a growth system related with a critical Ising model \cite{Dashti-Naserabadi19} comes as a surprise, since no such peculiarities are apparent in the system formulation. 

Notably, the values of the static scaling exponents and the type of scaling ansatz verified in Ref.\ \cite{Dashti-Naserabadi19} coincide with those found very recently for the tensionless KPZ equation (TKPZ), a particular case of the KPZ equation and thus a candidate continuum representative of the universality class exposed by the results in Ref.\ \cite{Dashti-Naserabadi19}. 
The TKPZ equation has attracted recent interest for 1D interfaces \cite{Cartes22,Rodriguez-Fernandez22,Rodriguez-Fernandez22b,Fontaine23}, not the least for its potential relevance for non-KPZ scaling behavior experimentally found in some KPZ-related systems \cite{Fontaine22,Wei22}. Actually, the 1D TKPZ equation has been shown to define a universality class of its own \cite{Rodriguez-Fernandez22,Rodriguez-Fernandez22b}, different from that of the standard KPZ equation, which encompasses scaling behavior earlier found for e.g.\ growth models related with isotropic percolation \cite{Asikainen02,Asikainen02b}. The TKPZ equation seems to be the first continuum model displaying intrinsically anomalous scaling \cite{Rodriguez-Fernandez22,Rodriguez-Fernandez22b} in absence of quenched noise or morphological instabilities. 

\begin{figure}
\begin{center}
\includegraphics[width=0.8\columnwidth]{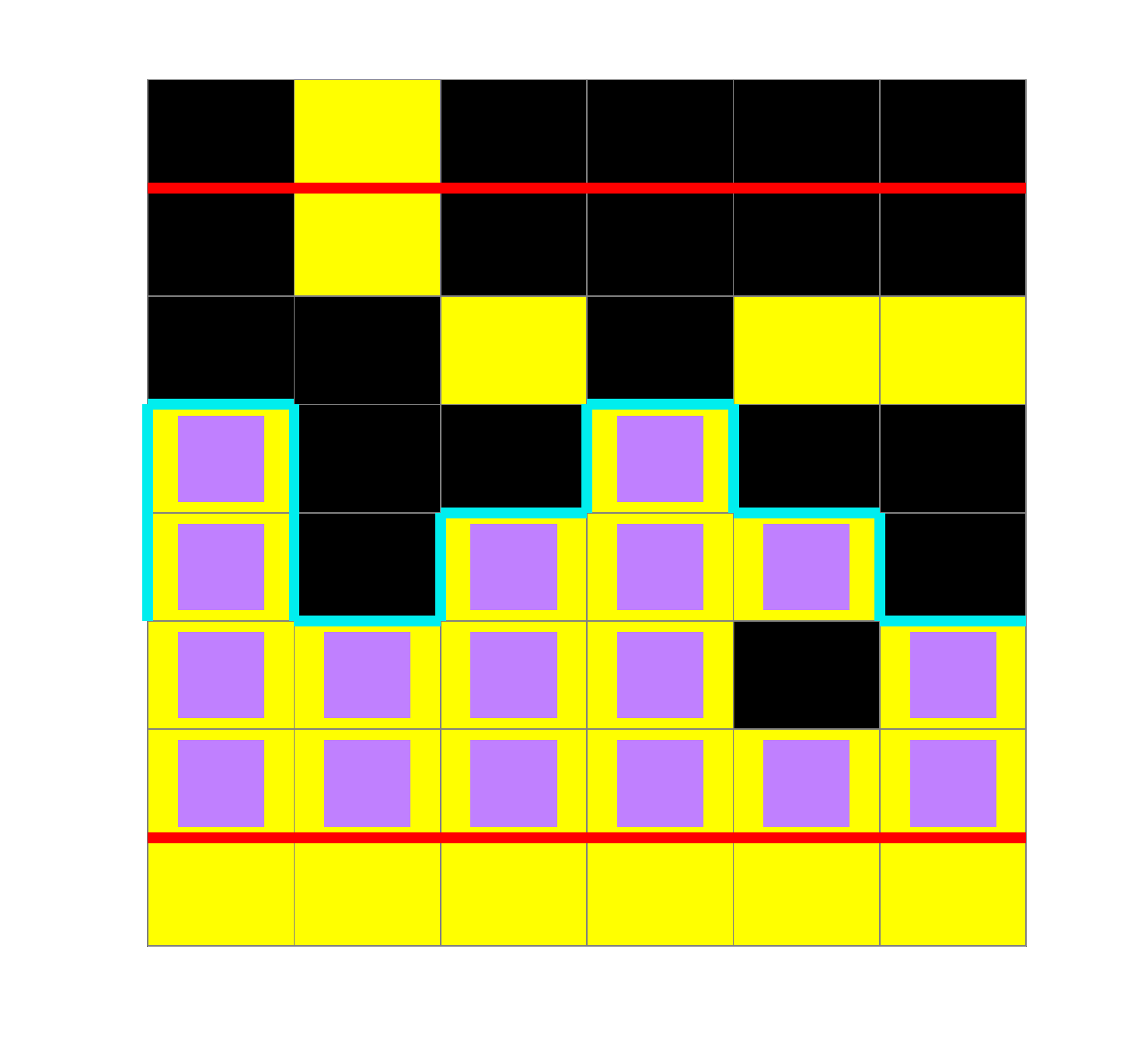}
\caption{Definition of the height profile $h(x)$ (cyan solid line) from a certain $6 \times 6$ Ising spin domain. See full definitions in Sec.\ \ref{sec:des} of the main text. Yellow and black cells correspond to $+1$ and $-1$ spin values, respectively. The spin rows below and above the bottom and top red lines illustrate the Dirichlet and Neumann boundary conditions, respectively. The (yellow) cells making up the cluster connected to the Dirichlet boundary via nearest neighbors are indicated by a purple bullet. The value of $h(x)$ corresponds to the height of the topmost one of these bulleted cells for each $x$.}
\label{C7fig:7}
\end{center}
\end{figure}

In this work, we revisit the interface growth model proposed in Ref.\ \cite{Dashti-Naserabadi19}. As the authors of that reference were addressing equilibrium properties of their system, specifically its classical roughening transition, no time-dependent behavior was reported and the kinetic roughening behavior was extracted from the system-size dependence of correlation functions. Thus, our goal is to study the time-dependent behavior of the system. 
Specifically, in our present work we study the full dynamics of the growth process defined in Ref.\ \cite{Dashti-Naserabadi19} at the critical temperature, $T=T_c$. We perform numerical simulations of the evolution of Ising spin domains in 2D and 3D, using both a Metropolis algorithm and an alternative coarse-grained approach based on the Ginzburg-Landau equation, using the boundary conditions proposed in Ref.\ \cite{Dashti-Naserabadi19} and described in Fig.\ \ref{C7fig:7}.

This paper is organized as follows. In section \ref{sec:obs}, we describe the observables we will measure to characterize the scaling of kinetic roughening processes. In section \ref{sec:des}, we describe the boundary conditions of the Ising model we will deal with, the definition of the interface, and the two different approaches that we have followed for the simulation of the time evolution of the system. Then, we describe the numerical results obtained both for 1D and 2D fronts (related to 2D and 3D Ising systems, respectively) via our Ginzburg-Landau approach in sections \ref{sec:1D} and \ref{sec:2D}. The results we have obtained via a Metropolis Monte Carlo approach are presented in the Appendix A. We discuss our results in section \ref{sec:disc}, followed by a summary and our conclusions in section \ref{sec:concl}.

\section{Observables\label{sec:obs}}

The observables which are going to be used in the characterization of the front dynamics are \textit{(i)} the global roughness, 
\begin{equation}
    W(t)= \Big\langle \sqrt{ {\frac{1}{L^d}} \int_{[0,L]^d} [h(\vec{x},t)-\bar{h}(t)]^2 d\vec{x} } \Big\rangle,
\end{equation}
where bar denotes spatial average, $L$ is the lateral system size, and brackets denote the average over different realizations of the noise, \textit{(ii)} the structure factor,
\begin{equation}
    S(\vec{k},t) = \langle |\hat{h}(\vec{k},t)|^2 \rangle,
\end{equation}
where $\hat{h}(\vec{k},t)$ is the Fourier transform of $h(\vec{x},t)$ and $\vec{k}$ is $d$-dimensional wave vector, and \textit{(iii)} the height-difference correlation function,
\begin{equation}
    G_2(\vec{r},t)= \sqrt{ \langle [\overline{h(\vec{x}+\vec{r},t)-h(\vec{x},t)}]^2 \rangle }.
\end{equation}

In a kinetic roughening process \cite{Barabasi95}, the global roughness scales with time as $W\sim t^{\beta}$ --- with $\beta$ being the so-called \textit{growth} exponent --- up to a saturation value $W_{sat} \sim L^{\alpha}$ which is reached at steady state. Here, $\alpha$ is the so-called \textit{roughness} exponent, which is related to the fractal dimension $D_f$ of the front considered as a self-affine fractal, as $\alpha=d+1-D_f$ \cite{Barabasi95}. The time $t_{sat}$ required for the system to reach the steady state scales as $t_{sat} \sim L^z$. Here, $z=\alpha/\beta$ is the so-called \textit{dynamic} exponent, which characterizes the time dependence of the lateral correlation length $\xi$ along the front as $\xi\sim t^{1/z}$ \cite{Barabasi95}. It is possible to define a scaling function $f_W$ that summarizes these scaling laws into the single expression \cite{Family85,Barabasi95}
\begin{equation}\label{C1eq:colW}
    W(L,t) \sim L^{\alpha} f_W \left( \frac{t}{L^z} \right),
\end{equation}
namely,
\begin{equation}
    f_W(u) \sim \left\{ \begin{array}{lcc}
             u^{\beta} & \mathrm{if} & u \ll 1, \\
             \mathrm{Constant} & \mathrm{if} & u \gg 1 .
             \end{array}
   \right.
\end{equation}
The local roughness measured over windows of size $l<L$ also scales with the window size $l$, with a local exponent $\alpha_{loc}$. Equivalently, the height-difference correlation function, Eq.\ \eqref{C1eq:colW}, scales as $G_2 \sim r^{\alpha_{loc}}$, where $r=|\vec{r}|$. In general, $\alpha_{loc}=\alpha$ (e.g.\ in the standard KPZ equation \cite{Barabasi95,Krug97}). However, there are some kinetic roughening systems in which the local and the global scalings of the roughness with the window size are different, i.e. $\alpha_{loc}\neq\alpha$. Such behavior is called \textit{anomalous scaling} or \textit{anomalous kinetic roughening} \cite{Schroeder93,DasSarma94,Lopez97,Ramasco00}. The structure factor scales in this case as \cite{Ramasco00}
\begin{equation}\label{C1eq:intrinsic}
    S(\vec{k},t) \sim k^{-(2\alpha + d)} f_{S} (k^z t),
\end{equation}
where $f_{S}$ behaves as
\begin{equation}\label{eq:scS}
    f_{S}(u) \sim \left\{ \begin{array}{lcc}
             u^{2 \alpha + d} & {\rm if} & u \ll 1 , \\
             u^{2(\alpha-\alpha_s)} & {\rm if} & u \gg 1.
             \end{array}
   \right.
\end{equation}
Here, $\alpha_s$ is an exponent conveniently measured in Fourier space which is equal to $\alpha_{loc}$ when $\alpha_{loc}<1$ \cite{Ramasco00}. For $\alpha=\alpha_{loc}$, the behavior described by Eqs.\ \eqref{C1eq:intrinsic} and \eqref{eq:scS} corresponds to the standard Family-Vicsek scaling ansatz \cite{Family85,Barabasi95}. As noted above, intrinsic anomalous scaling has been conjectured, via perturbative arguments, not to be in the asymptotic regime of continuum models which feature local interactions and time-dependent noise \cite{Lopez05}. 

Additionally, we will also check multiscaling behavior, where higher moments of the height-difference correlation function, namely,
\begin{equation}\label{C1eq:Gq}
    G_q(\vec{r},t)= \langle | h(\vec{x}+\vec{r},t)-h(\vec{x},t) |^q \rangle^{1/q} ,
\end{equation}
do not scale with the same roughness exponent for different values of $q$, i.e.\ for which $G_q(r)\sim r^{\alpha_q}$ with a $q$-dependent $\alpha_q$. In those cases, the morphologies are said to be multi-affine \cite{Barabasi95}. This kind of surface appears in e.g. surface growth models related to isotropic percolation \cite{Asikainen02,Asikainen02b}.

Finally, recent developments on kinetic roughening, mostly related to the KPZ equation (see e.g.\ Refs.\ \cite{Kriechebauer10,Corwin12,Halpin-Healy15,Takeuchi18}
and other therein), demonstrate the nontrivial role of the statistics of the height fluctuations to unambiguously identify the universality class, beyond scaling exponent values. Here, we will assess the field statistics by computing the probability distribution function (PDF) of the height fluctuations, as well as the time-dependent skewness,
\begin{equation}\label{eq:skew}
    {\cal S}(t)=\frac{1}{W^3(t)} \Big\langle {\frac{1}{L^d}} \int_{[0,L]^d} [h(\vec{x},t)-\bar{h}(t)]^3 d\vec{x} \Big\rangle ,
\end{equation}
and kurtosis,
\begin{equation}\label{eq:kurt}
    {\cal K}(t)=\frac{1}{W^4(t)} \Big\langle {\frac{1}{L^d}} \int_{[0,L]^d} [h(\vec{x},t)-\bar{h}(t)]^4 d\vec{x} \Big\rangle,
\end{equation}
where $W$ is the roughness of the $h(\vec{x},t)$ field.

\section{System description}\label{sec:des}

The physical system that we study in this work is in principle the same as that proposed in Ref.\ \cite{Dashti-Naserabadi19}. We define a 1D (resp.\ 2D) interface or front $h(x)$ (resp.\ $h(x,y)$) from a 2D (3D) spin domain $\{ s_{\vec{r}} \}$, where $s_{\vec{r}}=\pm 1$ are Ising spins and $\vec{r}$ takes values on a 2D (3D) square lattice of lateral size $L$. Dirichlet (fixed) and Neumann (free) boundary conditions are fixed on each boundary in the last coordinate of $\vec{r}$ (the ``vertical" or ``growth" one, $z$); specifically, $s_{\vec{r}}=+1$ (resp.\ $s_{\vec{r}'}=s_{\vec{r}}$) if $z=0$ (resp.\ if $z'=L+1$ and $z=L$). Periodic boundary conditions are considered in the other (transverse or substrate) dimensions. We will refer to these boundary conditions as \textit{magnet}. Then, a set of values 
$\mathcal{C}_{\vec{r}}$ is defined, such that $\mathcal{C}_{\vec{r}}=1$ if spin $s_{\vec{r}}$ is aligned with the $+1$ spins fixed at the Dirichlet bottom boundary, which are connected to each other via nearest-neighbor paths, and $\mathcal{C}_{\vec{r}}=0$ otherwise. For a 2D spin system, the height of the interface at a fixed substrate coordinate $x$ is finally defined as
\begin{equation}\label{C7eq:defh}
    h(x)=\max \{z \ | \ \mathcal{C}_{\vec{r}}=1 \},
\end{equation}
where $\vec{r}=(x,z)$. An illustrative 2D spin domain with $L=6$ under these boundary conditions, as well as its corresponding $h(x)$ interface profile, is depicted in Fig.\ \ref{C7fig:7}. An analogous procedure defines the height of the interface at a fixed substrate coordinate $(x,y)$ for a 3D spin system, where $\vec{r}=(x,y,z)$.

\begin{figure*}
    \centering
    \includegraphics{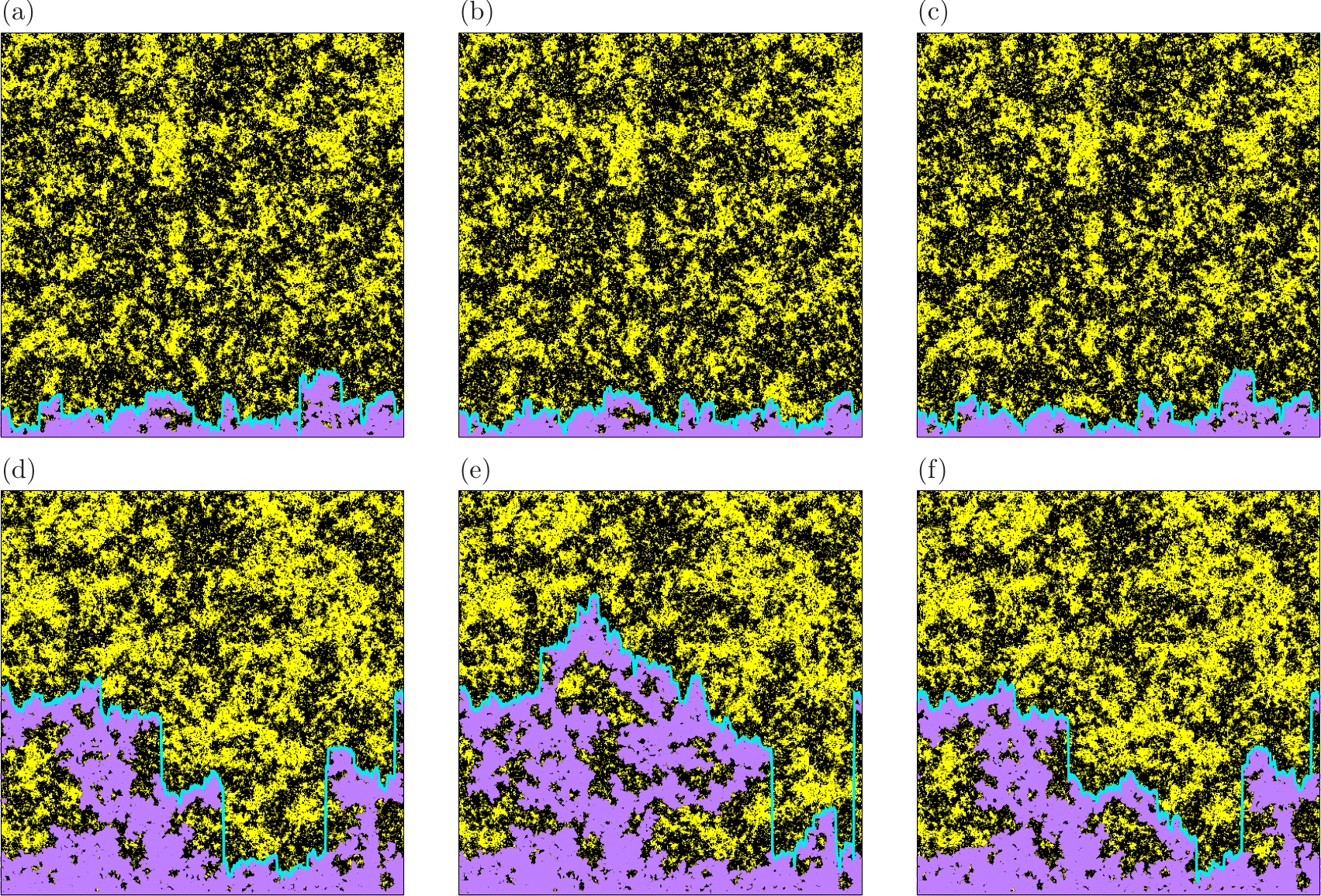}
    \caption{Snapshots of the 2D spin domain obtained from a numerical integration of Eq.\ \eqref{C7eq:GL}
    for $L=512$ and times $t=403.1,\, 403.2,\,\rm{and}\, 403.3$ for the top panels increasing from (a) to (c), and  $t=1099.3,\, 1099.4,\,\rm{and}\, 1099.5$ for the bottom panels increasing from (d) to (f). The same color coding of Fig.\ \ref{C7fig:7} is used here, with the only difference being that cells making up the cluster connected are now colored fully in purple. Again, for each time the front $h(x,t)$ is shown as a cyan solid line. This front performs relatively large vertical jumps at all times. Moreover, its overall shape can change quite abruptly in relatively short intervals as illustrated e.g.\ by the (d)$\to$(e)$\to$(f) sequence.}
    \label{fig:my_label}
\end{figure*}

\subsection{Metropolis approach}

The most straightforward method for the study of the dynamical evolution of the spin configurations of a ferromagnetic system consists in the use of Monte Carlo simulations \cite{Newman99}. A Metropolis algorithm can be used in order to simulate the full evolution of the spin field, while the equilibrium state of the model described in the previous paragraph was studied in Ref.\ \cite{Dashti-Naserabadi19} using Wolff's algorithm. For each Monte Carlo step, one random spin in a position $\vec{r}$ is chosen and flipped with probability $P_{\vec{r}}$, such that
\begin{equation}\label{C7eq:ProbMetropolis}
    P_{\vec{r}}= \left\{ \begin{array}{lcc}
             e^{-\Delta E_{\vec{r}}/k_B T} &   {\rm if}  & \Delta E_{\vec{r}} > 0, \\
             \\ 1/2 &  {\rm if} & \Delta E_{\vec{r}} = 0, \\
             \\ 1 &  {\rm if}  & \Delta E_{\vec{r}} < 0,
             \end{array}
   \right.
\end{equation}
where 
\begin{equation}\label{eq:hamil}
    H[\{ s_{\vec{r}} \}]= - J \sum_{\vec{r}} s_{\vec{r}} \big( \sum_{\vec{r}' \in \mathcal{N}(\vec{r})} \ s_{\vec{r}'} \big)
\end{equation} 
is the Ising Hamiltonian, $\mathcal{N}(\vec{r})$ is the set of all the nearest-neighbours for the position $\vec{r}$ on the 2D or 3D square lattice, $J>0$ is a ferromagnetic coupling, and $k_B$ is Boltzmann's constant. Hence, $\Delta E_{\vec{r}}= 2 J s_{\vec{r}}\left( \sum_{\vec{r}' \in \mathcal{N}(\vec{r})} \ s_{\vec{r}'} \right) $ is the energy change due to the spin flip at position $\vec{r}$. {The time scale is set to $t=N/L^d$, where $N$ is the number of Monte Carlo steps.}

This method yields a behavior in which boundary effects strongly affect the evolution of the front $h$, so that its scaling can not be unambiguously assessed for the system sizes we have been able to simulate. The details of these results are described in Appendix A. In view of this fact, we alternatively employ a coarse-grained approach that allows us to access the effective behavior of larger systems via the time-dependent Ginzburg-Landau equation.

\subsection{Ginzburg-Landau approach}

The (stochastic) Ginzburg-Landau (GL) equation \cite{Garcia-Ojalvo_Book}
\begin{equation}\label{C7eq:GL}
    \partial_t m = \frac{1}{2} \left( \nabla^2 m + m - m^3 \right) + \tilde{D} \eta,
\end{equation}
where $m(\vec{x},t)$ denotes the local magnetization field and $\eta$ is an uncorrelated white noise term, is an effective coarse-grained model, well-known to describe the evolution of the scalar magnetization of an Ising ferromagnet around thermal equilibrium \cite{Garcia-Ojalvo_Book,Taverniers14}. We use this model in order to simulate the full dynamic evolution of our Ising system. We also define here a coarse-grained spin lattice $\{ s_{\vec{x}} \}$ by discretizing $s_{\vec{x}}=+1$ if $m(\vec{x})>0$ and $s_{\vec{x}}=-1$ otherwise, from which we will define the field $h$ as in the original spin system, recall Eq.\ \eqref{C7eq:defh}. The same boundary conditions as those proposed in Ref.\ \cite{Dashti-Naserabadi19} are considered, see Fig.\ \ref{C7fig:7}.

Our purpose is to assess the behavior of this coarse-grained spin system at the noise amplitude {value $\tilde{D}=\tilde{D_c}$} corresponding to the Ising critical temperature $T_c$. {For such a value of the noise,} the relative fluctuation of the local magnetization field, 
\begin{equation}\label{eq:definicionM}
    M= \frac{\langle m^2(\vec{x}) \rangle- \langle m(\vec{x}) \rangle^2}{L^{d} {\tilde{D}}},
\end{equation}
exhibits a divergence at steady state as $M \sim L^{\gamma/\nu}$. Here, $\gamma=7/4$ and $\nu=1$ are the exact Ising critical exponents in two dimensions \cite{Garcia-Ojalvo_Book} and $\gamma\simeq 1.23$ and $\nu \simeq 0.63$ are the approximate values in three dimensions \cite{Cardy_book}.

Numerical simulations of Eq.\ \eqref{C7eq:GL} have been carried out in real space. A straightforward finite-difference scheme in space and an Euler scheme in time have been employed \cite{Garcia-Ojalvo_Book}, using $\Delta t=0.1$ and $\Delta x=1$. A homogeneous initial condition $m(\vec{x})=-1$ for all $\vec{x}$ except at the 
Dirichlet boundary, where $m=1$ at all times.

\section{Dynamics at $T=T_c$ for a one-dimensional interface}\label{sec:1D}

\subsection{$L\leq 2048$}

\begin{figure*}[!t]
\begin{center}
\includegraphics[width=2\columnwidth]{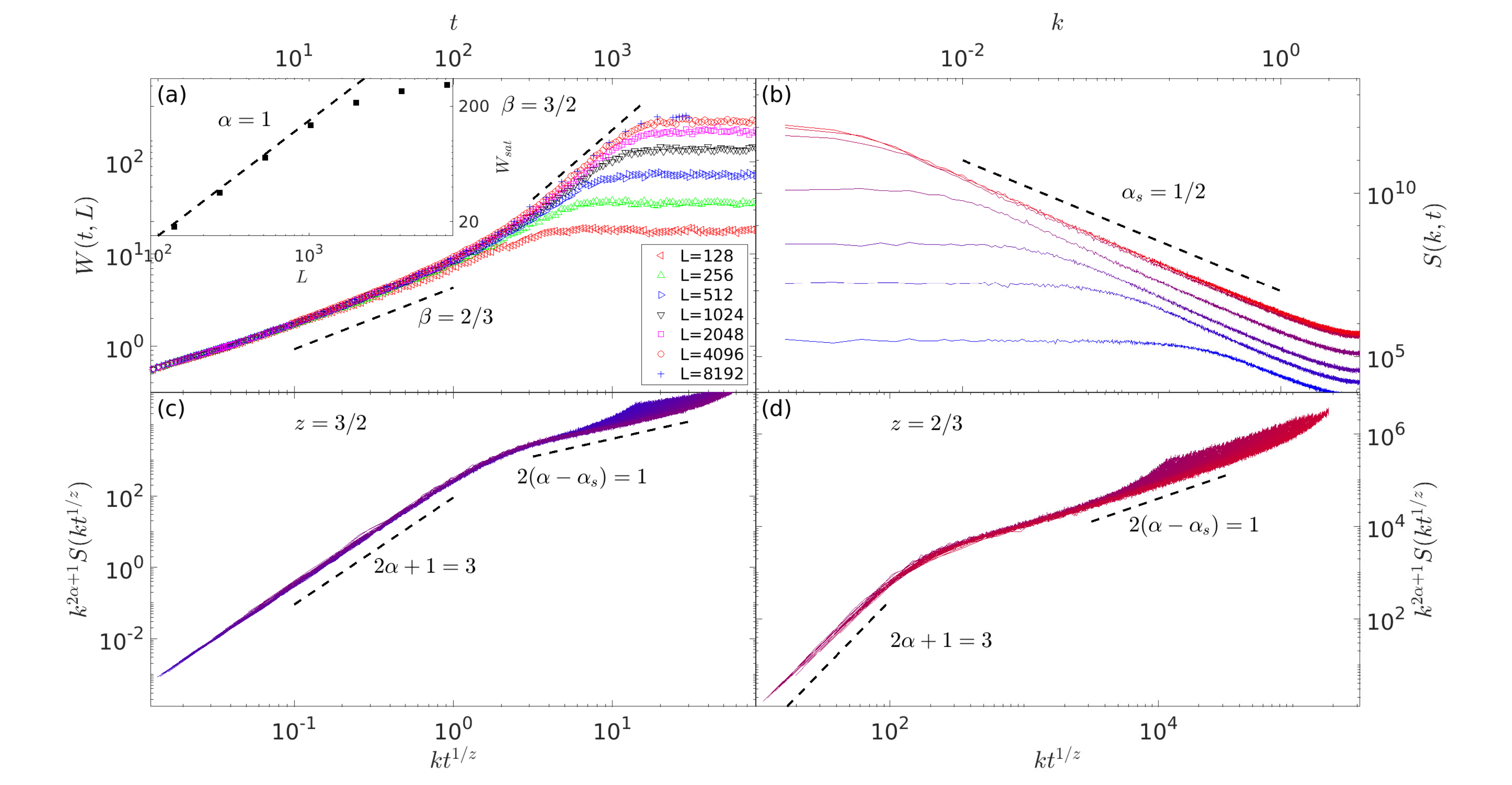}
\caption{ Time evolution for (a) the roughness $W(t,L)$ and (b) the structure factor $S(k,t)$ of the field $h(x,t)$ defined from the simulations of Eq.\ \eqref{C7eq:GL} for a 2D domain. The inset in (a) shows the scaling of the saturation roughness $W_{sat}$ with the system size $L$ for the values of $L$ given in the legend of the main panel. In (b) $L=8192$ and time values increase bottom to top. These $S(k,t)$ data are collapsed in (c) [resp.\ (d)] for early [late] times during which $z=3/2$ [$z=2/3$] and $L=2048$. Both collapses are for $\alpha=1$ and $\alpha_s=1/2$. In all panels, dashed lines correspond to the indicated values of the scaling exponents.}
\label{C7fig:9}
\end{center}
\end{figure*}

The entire critical dynamics of the field $h(x,t)$ is evaluated at $\tilde{D}=\tilde{D}_c$. In Appendix B, it is shown that $\tilde{D}_c\simeq 0.9$ corresponds to the critical temperature $T=T_c$ of the 2D Ising model not only with periodic boundary conditions \cite{Garcia-Ojalvo_Book} but also with the magnet boundary conditions shown in Fig.\ \ref{C7fig:7}. 

Figure \ref{fig:my_label} shows snapshots of the 2D spin domain obtained in a sample realization of our numerical simulations of Eq.\ \eqref{C7eq:GL}, which have been taken at increasing values of time from panel (a) through panel (f). The front $h(x,t)$ in each panel is shown as a cyan solid line. Such a line is seen to perform relatively many large vertical jumps. Its detailed shape can change quite dramatically in very short times when regions of the purple cluster connected to it by a few spins become ``pruned'' due to these spins flipping out of the cluster. This behavior is even more dramatically seen in the movie available in the supplemental material (SM) at Ref.\ \cite{sm} and already anticipates large values for the scaling exponents, see e.g.\ Ref.\ \cite{Barreales22} and other references therein.

\subsubsection{Scaling exponents}
\label{sec:expon}

Beyond surface morphologies, Fig.\ \ref{C7fig:9} shows the time evolution of both the surface roughness $W(t)$ and the structure factor $S(k,t)$, together with data collapses according to Eqs.\ \eqref{C1eq:intrinsic} and \eqref{eq:scS}. Note that Ref.\ \cite{Dashti-Naserabadi19} employed system sizes $L \leq 2048$, which, for the sake of comparison, are explicitly addressed in the present paragraph. The behavior that we obtain for larger values of $L \leq 8192$ is discussed in Sec.\ \ref{sec:largeL} below.

For the smallest system size considered in Fig.\ \ref{C7fig:9} (namely, $L=128$), the time increase of $W(t)$ is well characterized by a growth exponent value $\beta_1 \simeq 2/3$. Recalling that random deposition features $\beta_{RD}=1/2$ \cite{Barabasi95}, $\beta_1$ indicates very large interface fluctuations in time, as anticipated above. Furthermore, results for larger $L$ values indicate that this is a short-time behavior that is followed, for sufficiently large system sizes, by an even higher growth exponent $\beta_2 \simeq 3/2$. For intermediate $L$ and times, one might measure an intermediate value of $\beta$ close to that of the TKPZ equation ($\beta_{\rm TKPZ}=1$ \cite{Rodriguez-Fernandez22b}), but that is an apparent behavior, as borne out from the data collapses of the structure factor. Indeed, for $L=2048$ Fig.\ \ref{C7fig:9}(c) shows that early time $S(k,t)$ data collapse well using $\alpha\simeq1$ and $z_1\simeq3/2$ (hence, $\beta_1=\alpha/z_1=2/3$), while Fig.\ \ref{C7fig:9}(d) shows that long time data collapse well using $\alpha\simeq 1$ and $z_2\simeq2/3$ (hence, $\beta_2=\alpha/z_2=3/2$). The $\alpha\simeq1$ value is suggested by the system-size behavior of the roughness at steady state [see the inset of Fig.\ \ref{C7fig:9}(a) for $L\leq 2048$], while the spectral exponent value $\alpha_s\simeq1/2$ describes the $k$-dependent behavior of the structure factor, see Fig.\ \ref{C7fig:9}(b). Indeed, recall that, in the presence of intrinsic anomalous scaling, Eqs.\ \eqref{C1eq:intrinsic} and \eqref{eq:scS} imply that the structure factor scales as $S(k) \sim 1/k^{2\alpha_s+1}$ for $k\gg 1/t^{1/z}$ \cite{Ramasco00}. The intrinsically anomalous scaling ansatz which is verified and the roughness exponent values, $\alpha=1$ and $\alpha_s=1/2$, all coincide with those of the 1D TKPZ universality class \cite{Rodriguez-Fernandez22b}, as also obtained in the simulations of Ref.\ \cite{Dashti-Naserabadi19}. Note, in the latter reference this scaling behavior was obtained in equilibrium (at saturation, in our time-dependent approach), while we are presently assessing it along the time evolution of the system. However, in contrast with these TKPZ values for the roughness exponents, Fig.\ \ref{C7fig:9} rules out a ballistic  value for the dynamic exponent $z$ as in the TKPZ class, $z_{\rm TKPZ}=1$ \cite{Rodriguez-Fernandez22b}. For the sake of reference, the standard KPZ equation has a superdiffusive $z_{\rm KPZ}=3/2>1$ (as in our case for short times). Indeed, the value that we measure at long times, $z_{2}=2/3<1$, implies a spread of correlations that is even faster than ballistic, as occurs in the inviscid noisy Burgers equation \cite{Rodriguez-Fernandez22b}. It can also be found e.g.\ in suitable continuum models combining morphological instabilities with non-local interactions \cite{Nicoli09}.

\begin{figure}
\begin{center}
\includegraphics[width=\columnwidth]{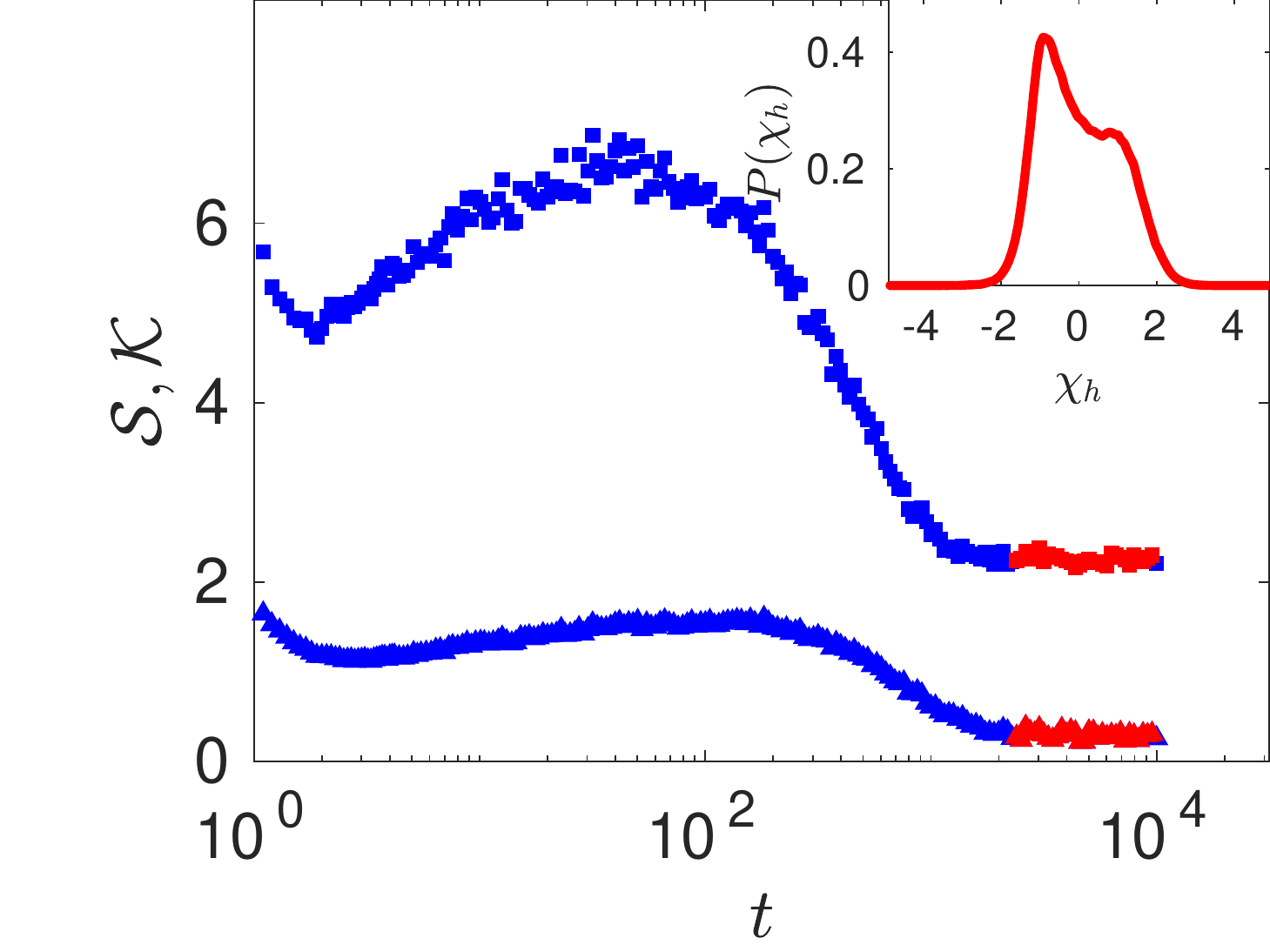}
\caption{Time evolution of the skewness (triangles) and kurtosis (squares) for the height fluctuations $\chi_h=(h-\bar{h})/{\rm std}(h)$ over 256 realizations from simulations of the GL model in 2D at critical noise $\tilde{D}= 0.9 \simeq \tilde{D}_c$ and $L=2048$. The inset shows the full height PDF for times which correspond to the steady state, as implied by the behavior of the skewness and kurtosis (red symbols).}
\label{C7fig:10}
\end{center}
\end{figure}

\subsubsection{Statistics of front fluctuations}

As noted above, beyond scaling exponent values the statistics of the height fluctuations is an additional trait of the system that identifies the kinetic roughening universality class. Hence, we next measure the fluctuations of the height field $h(x,t)$ at different times. Specifically, Fig.\ \ref{C7fig:10} shows our numerical results for the temporal evolution of the skewness and the kurtosis of the height fluctuations, and the full probability distribution function (PDF) of the steady-state fluctuations for our 1D fronts. Both ${\cal S}$ and ${\cal K}$ show very large values at short times which, after a nontrivial time evolution --- frequently found in continuous nonlinear models ---, eventually reach non-Gaussian (recall ${\cal S}_{\rm Gauss}=0$ and ${\cal K}_{\rm Gauss}=3$) values at steady state.

\begin{figure}
\begin{center}
\includegraphics[width=\columnwidth]{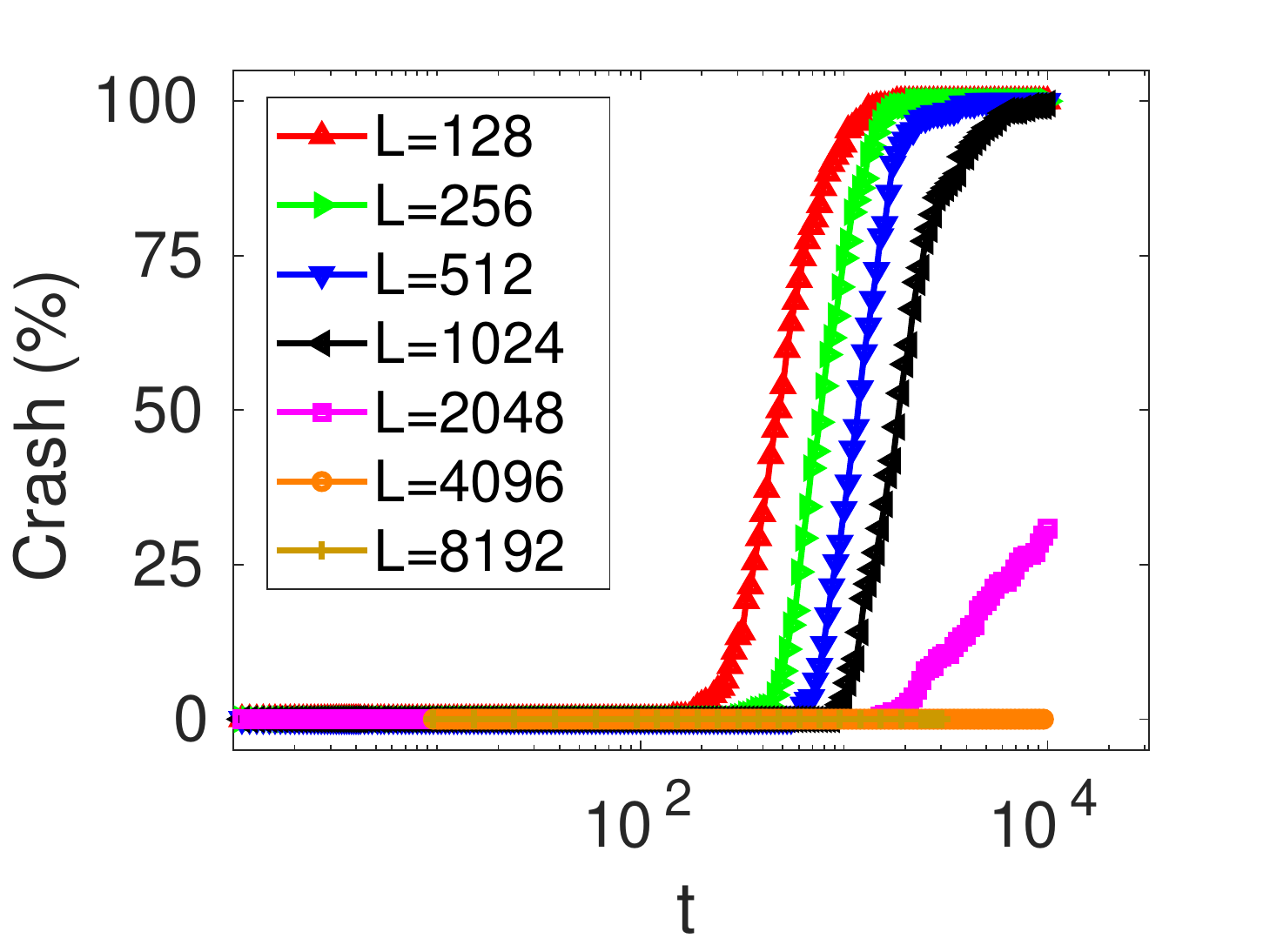}
\caption{Percentage of single realizations in which the maximum height $h$ has reached the system size $L$ (which we denote by Crash) at any previous time for different values of $L$. For larger values ($L=4196$ and $L=8392$) it is equal to zero for all $t$ (not shown).}
\label{C7fig:11}
\end{center}
\end{figure}

The full height PDF at saturation displayed in the inset of Fig.\ \ref{C7fig:10} is asymmetric indeed. Actually, we believe that the negative range of the fluctuations is over-represented, due to a distortion introduced by the Dirichlet boundary condition at the bottom of the 2D domain. To illustrate the interplay of the morphology with the boundary conditions, we show in Fig.\ \ref{C7fig:11} the time evolution of the mean value of $h$, as well as the percentage of the individual realizations for which some parts of the $h(x,t)$ front reach the top boundary, thus influencing the fluctuations, for different values of the system size $L$. We observe how this percentage decreases with the system size, reaching a negligible value for $L=2048$. Hence, the influence of the boundary in all measurements is expected to be very small for large system sizes.

\subsubsection{Multiscaling}

\begin{figure}
\begin{center}
\includegraphics[width=0.8\columnwidth]{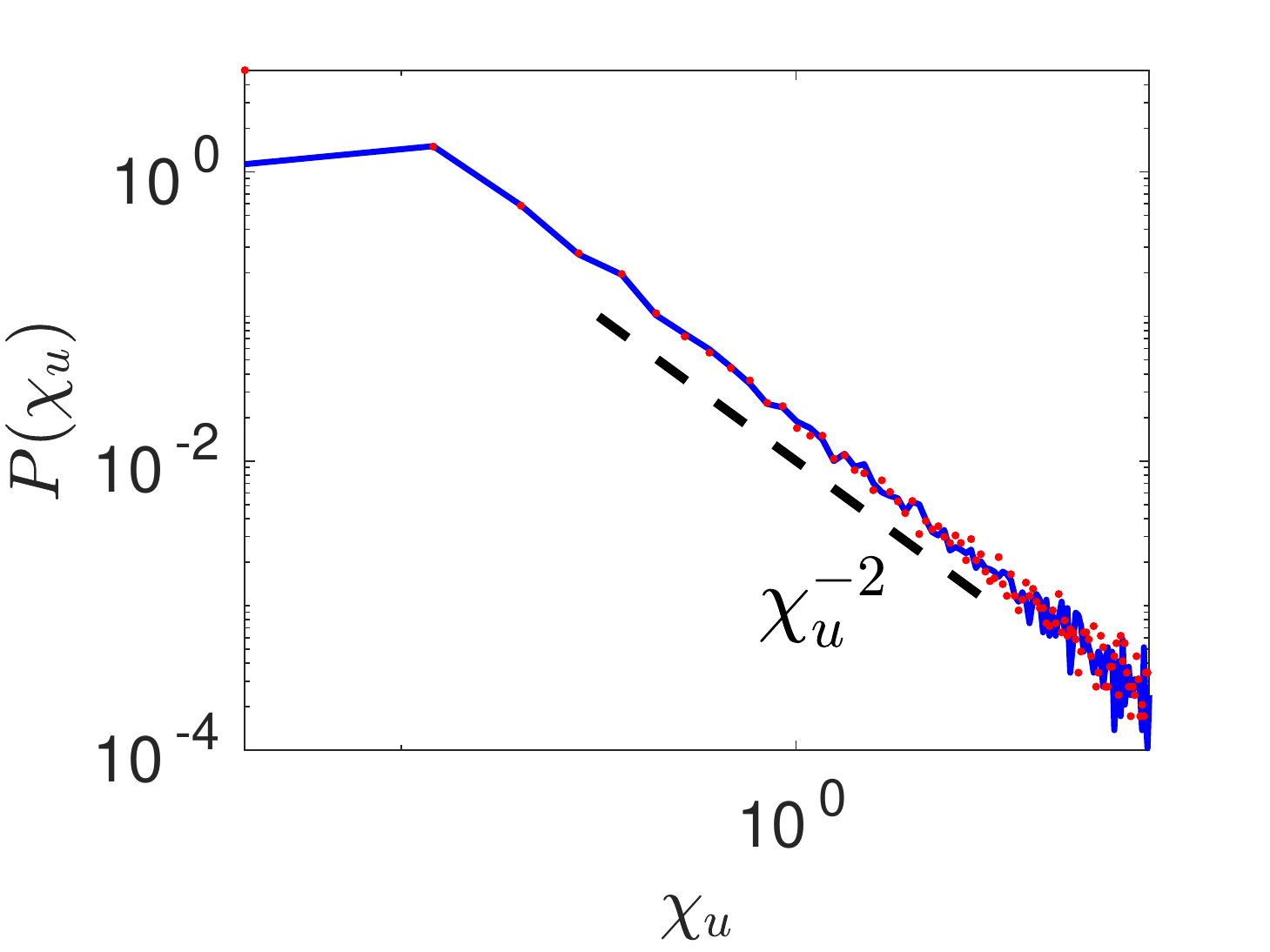}
\caption{ Histogram of slope values [$\chi_u=(u-\bar{u})/{\rm std}(u)$, where $u=\partial_x h$] from morphologies of the 2D GL growth model for $L=2048$ at steady state. The dashed line corresponds to the indicated power-law behavior with $\chi$. Red dots correspond to the negative slopes.}
\label{C7fig:Mul1}
\end{center}
\end{figure}

\begin{figure}
\begin{center}
\includegraphics[width=\columnwidth]{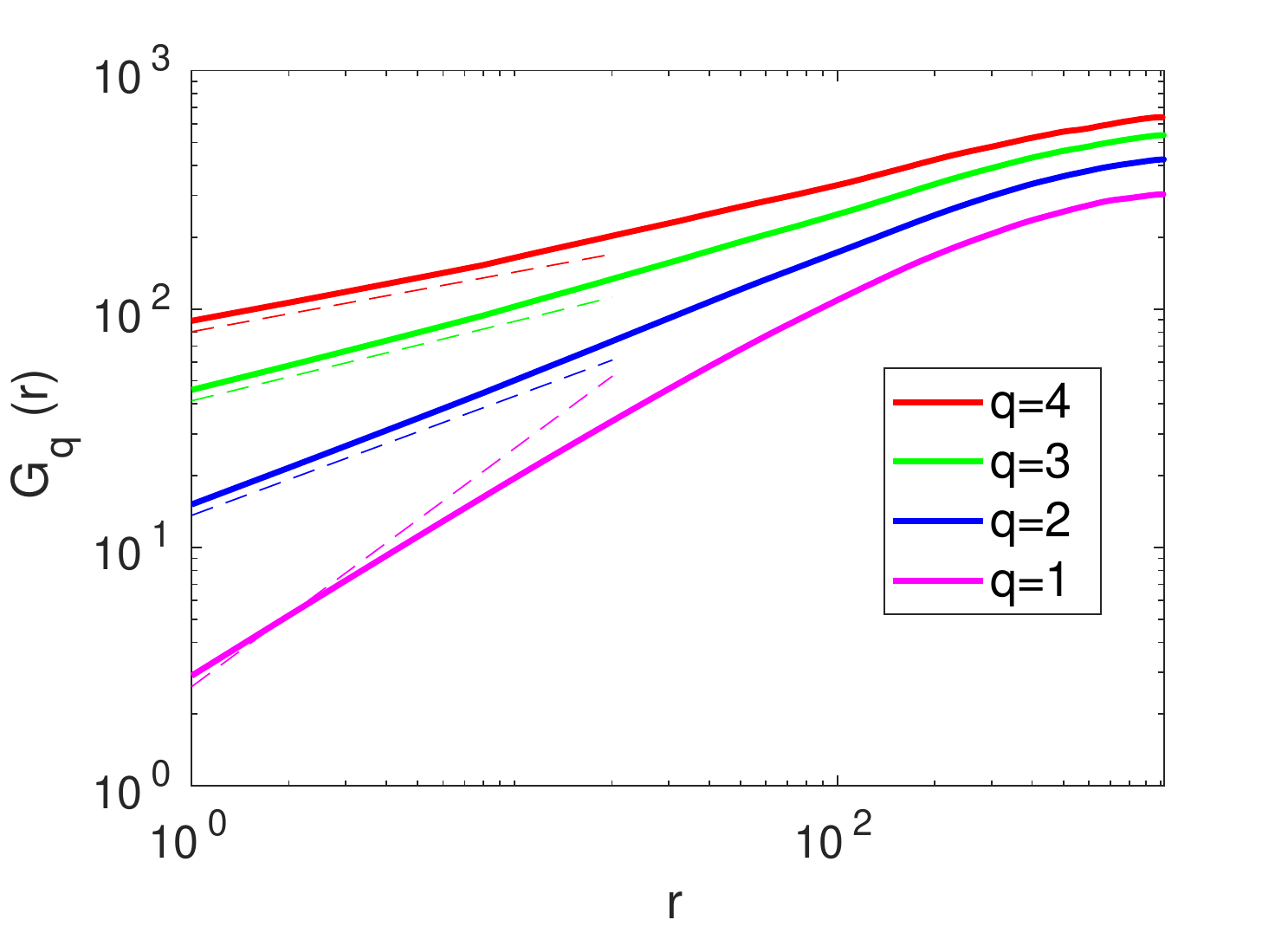}
\caption{First, second, third, and fourth moments (bottom to top) of the height-difference correlation function $G_q(r)$ from numerical simulations of the 2D GL model in the nonlinear growth regime for $L=2048$. Dashed straight lines correspond to the exact values of the slopes $\alpha_q=1/q$, as predicted in Refs.\ \cite{Asikainen02,Asikainen02b}.}
\label{C7fig:Mul2}
\end{center}
\end{figure}

The morphologies in the nonlinear growth regime from the 2D GL model exhibit an abundance of prominent slopes to the naked eye, recall Fig.\ \ref{C7fig:9}. In Fig.\ \ref{C7fig:Mul1} we assess the PDF of the slope field $u=\partial_x h$ for the GL interfaces. We observe in the figure that the tail of the PDF decays approximately as a power law ${P}(\chi_u) \sim \chi_u^{-2}$. In Ref.\ \cite{Asikainen02,Asikainen02b}, this type of slope statistics has been shown to imply multiscaling behavior, as different $q$-moments of the height-difference distribution, Eq.\ \eqref{C1eq:Gq}, were then shown not to scale with the same roughness exponent $\alpha$ for different values of $q$. Specifically, in Refs.\ \cite{Asikainen02,Asikainen02b}, a surface growth model related to isotropic percolation (invasion percolation without trapping) was studied numerically, finding that the statistics were well described by the power law ${P}(\chi_u) \sim \chi_u^{-2}$. In that case, a scaling analysis based on isotropic percolation implies that $\alpha_q=1/q$ for arbitrary $q$. This seems to also be the case for $q>1$ and not too large $r$ in our present numerical simulations, see Fig.\ \ref{C7fig:Mul2}.

\subsection{Larger values of $L$}
\label{sec:largeL}

As may have been noted by the reader when inspecting some of the figures presented in Sec.\ \ref{sec:expon}, a still different behavior is obtained for system sizes $L>2048$. Starting with Fig.\ \ref{C7fig:9}(a), the $L$-dependence of the saturation value of the roughness, $W_{sat}$, becomes weaker and becomes virtually inexistent for our largest system size, $L=8192$. A manifestation of this is the increasing deviation (for increasing $L$) of the effective $W_{sat} \sim L^{\alpha}$ law, from the $\alpha\approx 1$ value obtained for $L\leq 2048$, towards an effectively null roughness exponent for the largest $L$, implying the breakdown of kinetic roughening at large system sizes; see the inset of Fig.\ \ref{C7fig:9}(a). This interpretation is reinforced by the $S(k,t)$ data displayed for $L=8192$ in Fig.\ \ref{C7fig:9}(b). Indeed, in the presence of kinetic roughening (and both, under a FV scaling ansatz and under an intrinsically anomalous scaling ansatz), the value of $k_{\times} \sim 1/\xi(t)$ at which $S(k,t)$ changes from large-distance, uncorrelated, to short-distance, correlated behavior moves towards $k=0$ with increasing time. In contrast, the $S(k,t)$ curves in Fig.\ \ref{C7fig:9}(b) become time-independent for the longest times and fall one on top of the other. In this case, the (inverse of the) corresponding value $k^*_{\times}\sim \textrm{cnst.}$ signals a characteristic lateral length scale, instead of a time-dependent correlation length.

\section{Dynamics at $T=T_c$ for a two-dimensional interface}\label{sec:2D}

In this section we study the dynamics of the spin configurations in three spatial dimensions, which lead to the time evolution of two-dimensional fronts or interfaces $h(x,y,t)$. The noise amplitude that corresponds to the critical temperature in a 3D system is assessed again in analogy to our work for 2D spin domains. In this case, the critical $\tilde{D}$ takes the value $\tilde{D}_c \simeq 1.25$ \cite{Rodriguez-Fernandez22}.

\begin{figure}
\begin{center}
\includegraphics[width=\columnwidth]{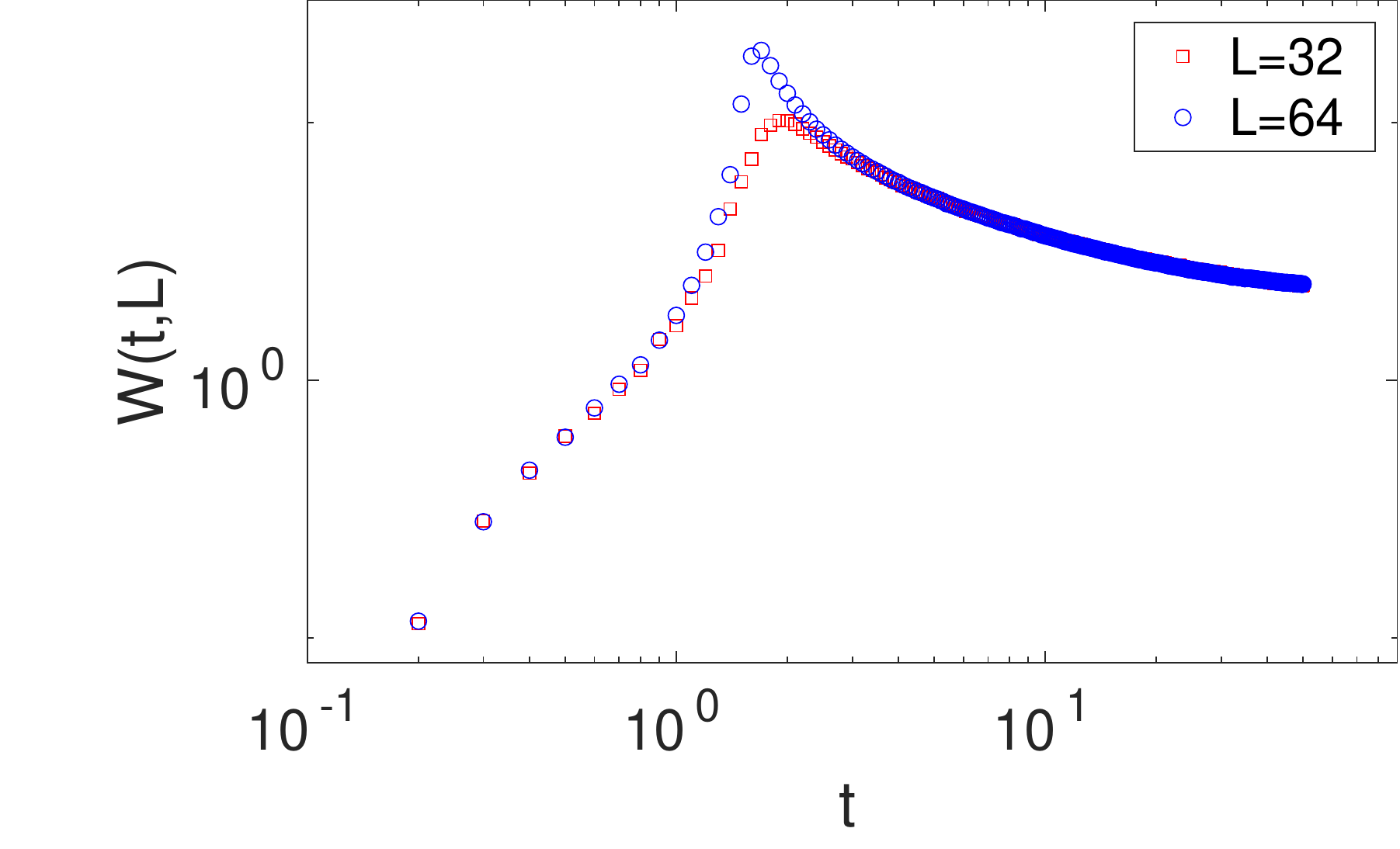}
\caption{Time evolution of the roughness $W$ for the height profiles obtained from numerical simulations of the GL model in three-dimensional domains, using boundary conditions as described in Fig.\ \ref{C7fig:7}, and for different values of the lateral system size $L$.}
\label{C7fig:13}
\end{center}
\end{figure}

\begin{figure}
\begin{center}
\includegraphics[width=\columnwidth]{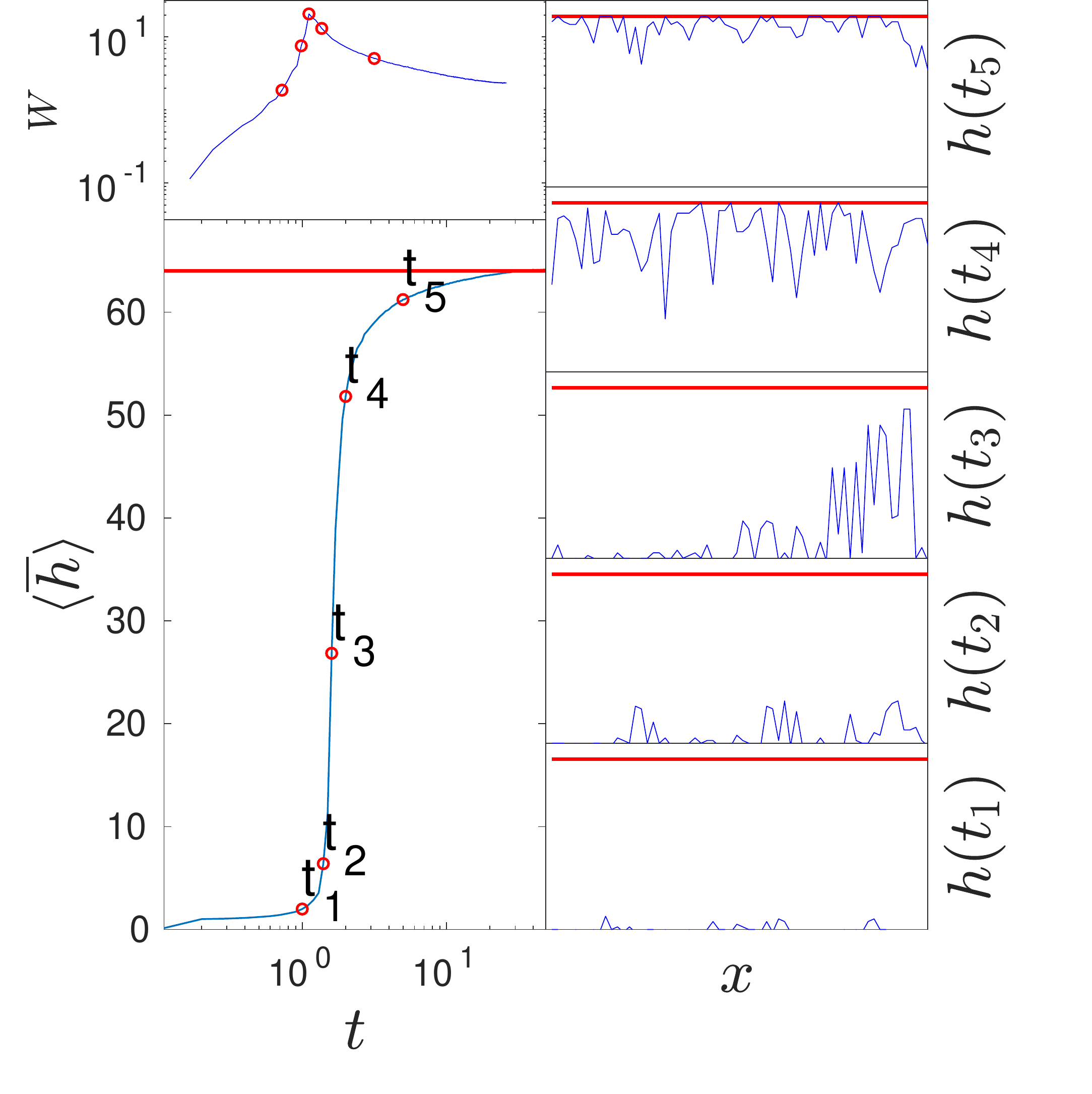}
\caption{Time evolution of the average mean height profile $ \langle \bar{h} \rangle $ obtained over 10 realizations of the noise from numerical simulations of the 3D GL model (left bottom panel) compared to the maximum achievable height (domain lateral size, horizontal red solid line). Left top panel shows the corresponding evolution of the surface roughness $W(t)$. One realization of a longitudinal cut $h(x,1,t)$ for each one of the highlighted times, $t_1$ to $t_4$ in the left panels, is shown in the right panels.}
\label{C7fig:14}
\end{center}
\end{figure}

In principle, our main interest lies with the kinetic roughening properties of the corresponding $h(x,y,t)$ interfaces. The evolution of the front roughness $W(t)$ is depicted for different system sizes in Fig.\ \ref{C7fig:13}. The growth of the roughness is suddenly interrupted and $W$ starts to decay for $t \gtrsim 20$. This behavior is induced by the upper boundary as we can appreciate in Fig.\ \ref{C7fig:14}. Indeed, the mean height $\overline{h}$ suddenly approaches the upper boundary at the same time in which the growth of the roughness is interrupted. For longer times the surface becomes pinned to this upper boundary, leading to the decrease in the roughness from that time on, much as it happens in our Metropolis approach both for 1D and 2D fronts (see Appendix A). With our computational power, we have not been able to find $L$ values which are free from this behavior. Moreover, both under the GL approach and under the Monte Carlo approach assessed in Appendix A, the peak value for $W(t)$ reached in the 2D interfaces increases with the system size. Hence, we deem it unlikely that these crahes can be avoided for even larger $L$ values than those considered here.

\section{Summary and Discussion \label{sec:disc}}

We have revisited the growth model formulated and studied in equilibrium in Ref.\ \cite{Dashti-Naserabadi19}, to explore its time-dependent properties, with the general goal to assess asymptotic intrinsic anomalous scaling in a system with time-dependent noise. More specifically, in view of the reported anomalous scaling ansatz and values of the roughness exponents \cite{Dashti-Naserabadi19}, a natural hypothesis to be checked was if the dynamical exponent also agreed with that of the TKPZ equation, making the model a member of its universality class. In such a case, moreover, the discrete growth model might provide a means to explore that universality class in higher dimensions ($d>1$), where the continuum TKPZ equation is conspicuously prone to instabilities \cite{Tabei04,Bahraminasab04,Rodriguez-Fernandez22}. In our simulations, we have rephrased the model in terms of a related stochastic, time-dependent Ginzburg-Landau equation.

For one-dimensional fronts, the model we are presently studying shows very strong fluctuations, related to the definition of the spin cluster which is linked with the spins located at the bottom boundary. Indeed, a few spin flips can suffice to attach/detach large spin subclusters, inducing strong variations in space and time for the front profile $h(x,t)$. In view of the size of these fluctuations, a natural source of concern is whether our scaling results are determined by a prompt interaction with the upper boundary that confines the system, but Fig.\ \ref{C7fig:11} indicates that this is not the case for (sufficiently) large systems including our largest values of $L$. For the smallest values of $L$, we demonstrate that the effect of the interaction with the upper boundary does not significantly perturb our results in Appendix C, were the full dynamics (data shown in Fig.\ \ref{C7fig:9}) is reanalyzed by discounting all the realizations in which such an interaction occurs.

Continuing with 1D fronts, we have found that their scaling behavior differs for $L\leq 2048$ (the range of $L$ values studied in Ref.\ \cite{Dashti-Naserabadi19}) and for larger values of $L$. In the former case, during the full time evolution of the system, our model does reproduce the expected intrinsic anomalous scaling behavior and the same numerical values of the roughness exponents ($\alpha=1$ and $\alpha_s=1/2$) of previous discrete (equilibrium) \cite{Dashti-Naserabadi19} and continuous \cite{Rodriguez-Fernandez22b} models. Actually, very similar kinetic roughening properties to these have been also measured \cite{Asikainen02,Asikainen02b} in another discrete model of invasion percolation. 
Moreover, the surfaces in that model were seen to display similar $P(\chi_u) \sim \chi_u^{-2}$ slope statistics and multiscaling properties to those we are reporting in Figs.\ \ref{C7fig:Mul1} and \ref{C7fig:Mul2}. However, this behavior for the statistics of surface slopes, being absent in the continuum system studied in Ref.\ \cite{Rodriguez-Fernandez22b}, is possibly related to the discrete nature of the surfaces and the single-valued approximation employed in the discrete models, as believed to be the case in other instances in the literature \cite{Castro1998,Castro2000,Nicoli2008,Castro2012,Marcos2022}. 

In spite of these similarities between the model we are studying and those in Refs.\ \cite{Rodriguez-Fernandez22b,Asikainen02,Asikainen02b}, the value of the dynamic exponent that we measure in our simulations crosses over in time between two different values, none of which coincide with the TKPZ value. Crossover behavior of the present type, in which both the short and long time behaviors are intrinsically anomalous and only differ by the value of $z$ seems quite rare in the literature. Moreover, and as noted in Sec.\ \ref{sec:intro}, the fact that the asymptotic behavior remains intrinsically anomalous for a system with time-dependent noise contradicts theoretical expectations \cite{Lopez05} based on (perturbative) renormalization group arguments for continuum models.

Related with the latter fact, our simulations of 1D fronts show yet a different behavior for $L>2048$, which implies that the scaling found for smaller systems is not asymptotic. Indeed, as suggested by the behavior of the roughness $W(t)$ and the surface structure factor $S(k,t)$, kinetic roughening seems to break down at sufficiently large distances and times. Very analogous behavior has been recently reported for other kinetically rough systems \cite{Barreales22}. Thus for instance, an off-lattice generalization of the ballistic deposition model \cite{Dias14}, which has been shown to describe quite accurately \cite{Dias18} kinetically rough 1D fronts emerging in coffee-ring formation processes in colloidal systems \cite{Yunker13,Nicoli13,Yunker13b,Oliveira14}, has been shown to display effective intrinsically anomalous scaling at small to intermediate scales; however, such a scaling behavior breaks down at the largest time and length scales, being overridden by some sort of morphological instability inducing the formation of macroscopic shapes \cite{Barreales22}. The intrinsic anomalous scaling becomes problematic in the limit $L\to\infty$ as $S(k,L)\sim L^{2(\alpha-\alpha_s)}$ hence it is natural to expect it to be not an asymptotic behavior but a transient in most real systems.

In the case of 2D fronts in our 3D Ising systems, we obtain a very fast increase of the surface roughness due to the fast evolution of the interface height, in such a way that no non-trivial scaling behavior develops. This is the case both, for our Ginzburg-Landau and our Metropolis approaches. Here, note that while Ref.\ \cite{Dashti-Naserabadi19} reports a 3D surface roughness that does scale (at equilibrium) with system size under analogous conditions, it does so as if the system was above the upper critical dimension, which is unexpected taking into account that 2D sections of that same system behave as the 1D interfaces of 2D Ising systems.

Considering the detailed dynamical behavior that we obtain, the $\beta>0.5$ exponent values observed for our 1D fronts at small to intermediate scales and the large-scale breakdown of kinetic roughening might be anticipating the singular behavior that arises in higher dimensions. Recall that larger values of $\beta$ are usually taken as indicative of morphological instabilities for systems with time-dependent noise, as $\beta=1/2$ characterizes the roughness increase for the random deposition process \cite{Barabasi95}.

The description of the Ising system via the GL equation seems feasible in the one-dimensional case and has been extended to the study of the evolution of two-dimensional interfaces. However, the interaction of the interface with the free boundary of the system does not allow us to follow the long-time evolution in these 2D substrates. This is the case both for the GL and also for the discrete approach to the dynamics based on {our} Metropolis algorithm. While we are able to avoid the crash between the interface and the upper boundary in the simulations of the 1D fronts by scaling up the effective system size through a coarse-grained approach, this seems not to be possible for the 2D fronts. Indeed, in the latter case, this phenomenon also occurs even under the coarse-grained GL approach, {with the} peak values for $W(t)$ {increasing} with the system size under both approaches.

\section{Conclusions}\label{sec:concl}

In conclusion, we have shown that the 1D interfaces defined over the Ising spin systems as in Ref.\ \cite{Dashti-Naserabadi19} do not feature asymptotic kinetc roughening for large systems sizes ($L>2048$). For the previously explored range of $L$ values ($L\leq 2048$), we do reproduce the intrinsically anomalous scaling Ansatz and the values of the global and local roughness exponents obtained for the equilibrium discrete Ising model \cite{Dashti-Naserabadi19} and for some continuous systems \cite{Rodriguez-Fernandez22b}. In our present case, we have employed a Ginzburg-Landau approach to the dynamic behavior of the Ising system which provides better results than those obtained from a simple Metropolis algorithm. Still, the dynamic exponent crosses over in our simulations between short and long-time values, none of which agree with that of the TKPZ continuous model. The intrinsic anomalous scaling found for $L\leq 2048$ is perhaps related to non-localities in the interactions at play, e.g.\ the effect that single spin flips have in the dynamics of large regions of the front. In these cases the interface shows strong fluctuations with multiscaling and fat-tailed slope statistics --- perhaps related to the discreteness of the spins and the single-valued approximation of the fronts---, all of which (except again for the value of $z$) agree with earlier discrete models of invasion percolation.

On the other hand, the 2D fronts obtained through our approach present a “discrete blow-up”, which might be analogous to that found in the 2D tensionless KPZ equation \cite{Tabei04,Bahraminasab04,Rodriguez-Fernandez22b}. In our 3D spin model, this “blow-up” is due to the interaction with the system boundaries and is reminiscent of analogous behavior we have found for the 1D fronts of our smallest 2D systems. To extract more definitive conclusions on the behavior of the 2D fronts (including the analysis of their symmetries \cite{Dos_Anjos21}, provided the universality class can be defined), improved simulations of our non-equilibrium system seem required which access substantially larger system sizes, perhaps via cluster algorithms akin to those employed at equilibrium in Ref.\ \cite{Dashti-Naserabadi19}. 

\section{Acknowledgments}
We thank J.\ M.\ L\'opez for discussions and suggestions. Part of our numerical simulations were done in Uranus, a supercomputer cluster located at Universidad Carlos III de Madrid and funded jointly by EU-FEDER and the Spanish Government via Grants No.\ UNC313-4E-2361, No.\ ENE2009-12213-C03-03, No.\ ENE2012-33219, and No.\ ENE2015-68265, and via Grant No.\ SIMTURB-CM-UC3M from the Convenio Plurianual of Comunidad de Madrid (CAM, Spain).

In addition, this work has been partially supported by Ministerio de Ciencia e Innovaci\'on (Spain), by Agencia Estatal de Investigaci\'on (AEI, Spain, 10.13039/501100011033), and by European Regional Development Fund (ERDF, A way of making Europe) through Grants No.\ PGC2018-094763-B-I00, No.\ PID2022-140217NB-I00, and No.\ PID2021-123969NB-I00, and by CAM (Spain) under the Multiannual Agreements with UC3M in the line of Excellence of University Professors (EPUC3M14 and EPUC3M23), in the context of the V Plan Regional de Investigaci\'on Cient\'{\i}fica e Innovaci\'on Tecnol\'ogica (PRICIT). E.\ R.-F.\ acknowledges financial support from CAM through contract No.\ 2022/018 under the EPUC3M23 line and from Universidad Carlos III de Madrid through the Margarita Salas program.

\appendix

\section{Simulation results using the Metropolis algorithm}

The evolution of the interface field $h$ defined as described in Section \ref{sec:des}, i.e.\ using Eqs.\ \eqref{C7eq:defh} and \eqref{C7eq:ProbMetropolis} for both, 2D and 3D spin lattices, has been measured at $T=T_c$, where $T_c=2/\ln{(1+\sqrt{2})}\simeq 0.44$ is the exact value for the 2D square lattice and $T_c\simeq 0.22$ for the 3D cubic lattice \cite{Dashti-Naserabadi19}.

\begin{figure}
\begin{center}
\includegraphics[width=\columnwidth]{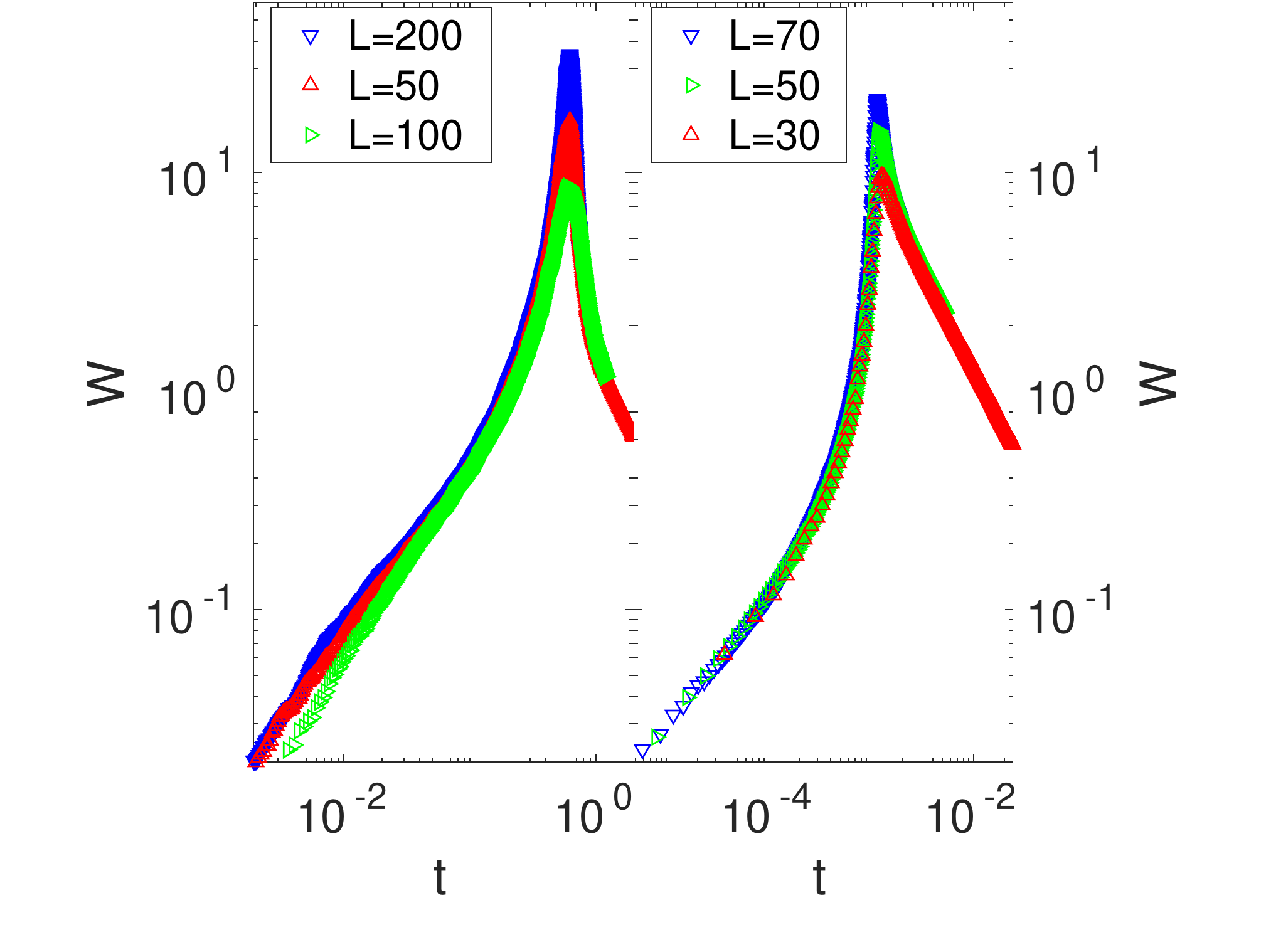}
\caption{Time evolution of the roughness $W$ for the height profiles obtained in the Metropolis evolution of an Ising system in both, two and three-dimensional domains (left and right panels, respectively) with boundary conditions as described in Fig.\ \ref{C7fig:7} and for different values of the lateral system size $L$.}
\label{C7fig:15}
\end{center}
\end{figure}

We find very fast growth of the roughness $W$ with time, as shown in Fig.\ \ref{C7fig:15} for different lateral system sizes $L$ for both 2D and 3D spin domains, hence 1D and 2D fronts. Such a fast growth process is interrupted when the mean height approaches the boundary of the system, leading to an abrupt decrease in the roughness from that time on. This behavior is very similar to that found in our Ginzburg-Landau approach for 3D domains. As the GL equation provides a coarse-grained description involving continuum, instead of discrete, values of the local degrees of freedom, it might be describing effectively larger system sizes with a comparable computational cost.

\section{Identification of the critical temperature}

\begin{figure}[!h]
\begin{center}
\includegraphics[width=\columnwidth]{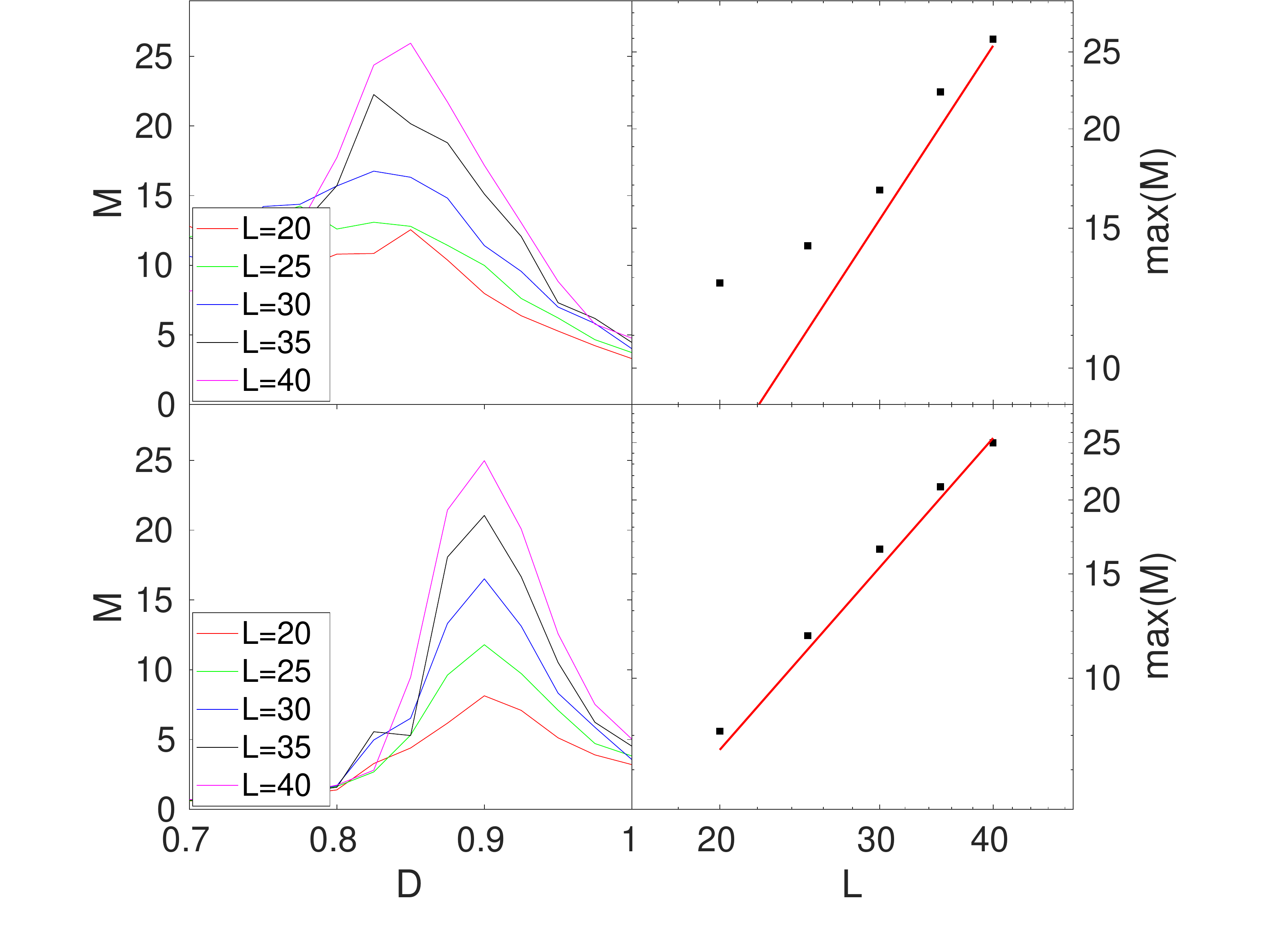}
\caption{Determination of the noise strength $D_c$ corresponding to the critical temperature $T_c$ for the GL equation in a 2-dimensional spatial domain. Left panels show the magnetization fluctuation $M$ at steady state for different values of $D$ and lateral system size $L$ [magnet boundary conditions (top) and periodic boundary conditions (bottom)]. At $D=D_c$, corresponding to $T=T_c$, $M$ diverges with $L$ as a $M\sim L^{7/4}$ (red solid line) as expected from the 2D Ising critical exponents $\gamma=7/4$ and $\nu=1$ \cite{Cardy_book}.}
\label{C7fig:8}
\end{center}
\end{figure}

We assess the behavior of the Ginzburg-Landau equation, Eq.\ \eqref{C7eq:GL}, at different values of the noise strength $D$ in order to determine the noise amplitude corresponding to the critical temperature $T_c$. In Fig.\ \ref{C7fig:8} we show how the relative fluctuation of the magnetization field --- see Eq.\ \eqref{eq:definicionM} --- at steady state $t \gg 1$ exhibits a divergence as $M \sim L^{\gamma/\nu}$ for the critical value $D=D_c\simeq 0.9$ corresponding to the critical temperature $T=T_c$. Here, $\gamma=7/4$ and $\nu=1$ are the Ising critical exponents in two dimensions \cite{Garcia-Ojalvo_Book}. This divergence is more clear if we consider a spin system in which all the boundary conditions are periodic, as shown in the bottom panels of Fig.\ \ref{C7fig:8}.

\section{Dynamics without those realizations that interact with the upper boundary}

\begin{figure*}[!h]
\begin{center}
\includegraphics[width=2\columnwidth]{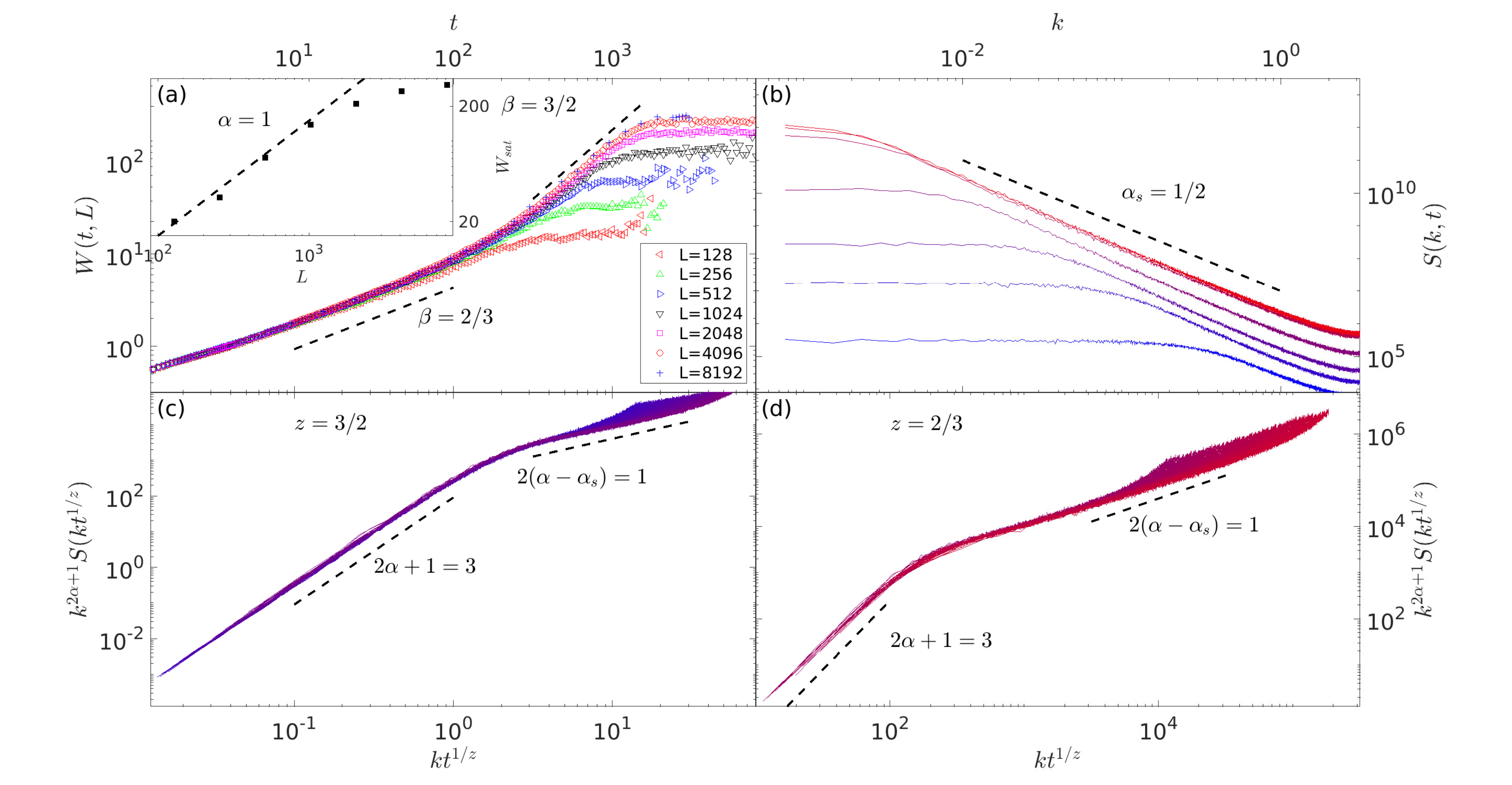}
\caption{Same as Fig.\ \ref{C7fig:9}, but only using $h(x,t)$ data in which the fronts does not reach the upper system boundary at any point, time, or realization.
}
\label{fig:remade}
\end{center}
\end{figure*}

In Fig.\ \ref{fig:remade}, the results of Fig.\ \ref{C7fig:9} are shown but discarding all realizations for which the surface profile touches the upper boundary at any given time.

\end{document}